\documentclass[11pt]{article}
\begin{document}
\title
{Noncommutative quantum mechanics of simple matter systems interacting with circularly polarized gravitational waves}
\author{
{\bf {\normalsize Sunandan Gangopadhyay}$^{a,c}
$\thanks{sunandan.gangopadhyay@gmail.com}},
{\bf {\normalsize Anirban Saha}
$^{a,c}$\thanks{anirban@iucaa.ernet.in}},
{\bf {\normalsize Swarup Saha}$^{b,a}$\thanks{saha18swarup@gmail.com}}\\[0.2cm]
$^{a}$ {\normalsize Department of Physics, West Bengal State University,
Barasat, Kolkata 700126, India}\\[0.2cm]
$^{b}$ {\normalsize Department of Radiotherapy and Nuclear Medicine, Barasat Cancer Research and Welfare Center,}\\
{\normalsize Barasat, India}\\[0.2cm]
$^{c}${\normalsize Visiting Associate in Inter University Centre for Astronomy $\&$ Astrophysics,}\\
{\normalsize Pune, India}\\[0.3cm]
}
\date{}

\maketitle

\begin{abstract}
\noindent The response of a test particle, both for the free case and under the harmonic oscillator potential, to circularly polarized gravitational waves 
is investigated in a noncommutative quantum mechanical setting. The system is quantized following the prescription in \cite{ncgw1}. Standard algebraic techniques are then employed to solve the Hamiltonian of the system. The solutions, in both cases, show signatures of the coordinate noncommutativity. In the harmonic oscillator case, this signature plays a key role in altering the resonance point and the oscillation frequency of the system.  
\end{abstract}

\maketitle
\section{Introduction}

\noindent It is now well understood that astrophysical objects like binary pulsars, quasars, giant black holes in galactic centres and high energetic events like gamma ray bursts emit gravitational waves (GWs). Since the indirect discovery of their existence from the orbital decay of binary pulsar PSRB 1913+16 \cite{HT} GWs have been at the focus of intense research both at theoretical as well as experimental fronts. 
Predicted originally by Einstein's general relativity, GWs are tiny disturbances in the fabric of spacetime which makes the proper length between two events oscillate. Based on this oscillation, the GWs are decomposed into two independent polarization modes, namely the linear and circular polarization \cite{Magg}. 

In particular, the circular polarization gives a way to describe whether the background has asymmetry with respect to magnitudes of right-handed and left-handed waves \cite{kah, seto}. Interestingly, the most recent BICEP results \cite{BICEP} claim to have detected the B-type polarization mode in the cosmic microwave background (CMB) which is caused by the left/right circular polarized GWs. This detection, though indirect in nature, has reinforced the relevance of circularly polarized GWs. Also the PLANCK data \cite{Planck1, Planck2} is currently being analyzed for similar results. 
From the phenomenological point of view various string-theory motivated modified theories of gravity also predicts circularly polarized GW. For example, Chern-Simons gravity, together with a Gauss-Bonnet term results in a super-inflationary phase in cosmological evolution \cite{mod_1, mod_2, mod_3, mod_4, mod_5} during which an instability in gravitational wave modes should exist \cite{mod_6, mod_7, mod_8} which generates the circular polarization of primordial GW. Thus the existence or otherwise of Chern-Simons gravity can also be probed by looking for the circular polarization mode. 

Various ground/space based interferometric GW detectors have been set up in the last couple of decades for direct detection of GWs. They basically involve the measurement of the relative optical phase shift between the light paths in two perpendicular km-length arm cavities. This phase shift is caused by the tiny displacement of two test mass mirrors hung at the end of each cavity, induced by passing GWs. The typical amplitude of GWs $h \sim \frac{\delta L}{L}$ emitted by binary systems in the VIRGO cluster of galaxies, at a distance 20 Mpc at 100 Hz, which the various GW detecters like \cite{abramovici}-\cite{lisa} are designed to probe, is $\sim 10^{-21}$. Since the cavity arm length $L \sim 1 {\rm km}$, $\delta L$ will be $\sim 10^{-18}{\rm m}$. This suggests that experimental evidence for the GWs is likely to appear at the quantum mechanical level\cite{Caves}. Besides, it is long believed that GWs would play a key role in understanding the interplay of classical and quantum gravity at a quantum mechanical level. 

A well-known way to impliment quantum gravity effects in a low-energy regime is to impose a noncommutative geometry among the spatial coordinates \cite{Dop1, Dop2, Alu, snyder} and construct the corresponding NC quantum mechanics. In this spirit, we have investigated the quantum dynamics of a free test particle \cite{ncgw1} and a harmonic oscillator \cite{ncgw2} in noncommutative (NC) space under the influence of linearly polarized GWs earlier. The motivation for carrying out such analysis stems from the fact that in the formulations of NC quantum mechanics \cite{mezin}-\cite{gan} and NC quantum field theory \cite{szabo}-\cite{sun}, the upper bound on NC length-scales \cite{carol, RB, ani1, stern, ani2} turns out to be comparable to the $\delta L$ arising in the context of GWs. Therefore the effect of NC space geometry may be a potential noise source in GW detectors. Our formulation in \cite{ncgw1, ncgw2} showed some interesting NC effects croping up in the GW-matter interaction. There we have specifically used the linear polarization mode of the GW to keep the computation simple. Note that if the BICEP data is corroborated by the soon-to-arrive PLANCK results the search for direct detection of circularly polarized GWs will surely gain impetus. Therefore it is imparative that we extend our earlier studies \cite{ncgw1, ncgw2} for the case of circularly polarized GWs. In the present paper we do the same. Note that due to rotation of the GW polarization vectors in the present case, the method of analysis presented in \cite{ncgw1, ncgw2} has to be modified in a non-trivial way. Therefore, we first formulate the case for a free particle, check the cocsistency of our result in the appropriate limit and only then move on to deal with the more involved harmonic oscillator case. For the latter, a new resonance effect, different in nature from the linear polarization case analyzed earlier \cite{ncgw2} is found in our calculation.

The paper is organized as follows. In section 2, we present the detailed methodology for constructing the NC quantum mechanics of simple matter system interacting with circularly polarized GWs, taking the example of a free particle. 
In section 3, we use this methodology to analyse the quantum dynamics of the harmonic oscillator in a similar setting and discuss the results. Finally, we conclude in section 4.


\section{Free particle in noncommutative space interacting with circularly polarized gravitational wave}
\noindent We start with  the example of a free particle in NC space interacting with circularly polarized GWs. The Hamiltonian of a scalar spin zero particle in the presence of a passing GW in transverse traceless (TT) gauge reads \cite{speli}
\begin{equation}
{H} = \frac{1}{2m}\left({p}_{j} + m \Gamma^j_{0k} {x}^{k}\right)^2 
\label{e90}
\end{equation}
where ${x}^{j}$ is the proper distance of the particle from the origin and $m$ is its mass\footnote{We assume that the GW detectors are reasonaby isolated so that external forces other than the GW interaction are negligible.}.

\noindent The equation of motion for ${x}^{j}$ which follows from this Hamiltonian is the well known geodesic deviation equation in the proper detector frame \cite{Magg} 
\begin{equation}
m\frac{d^2 {x}^{j}}{dt^2} = - m{R^j}_{0,k0} {x}^{k}.  
\label{e55}
\end{equation}
Eq.(\ref{e55}) describes the time evolution of the proper distance in TT-gauge frame as long as the spatial velocities involved are non-relativistic. Also, $|{x}^{j}|$ has to be much smaller than the typical scale over which the gravitational field changes substantially. This situation is referred to as the \textit{small velocity and long wavelength limit}.

\noindent Considering the GWs to be propagating along the $z$-axis, one can essentially focus on the motion of the particle in the 
$(x, y)$ plane since ${\Gamma^j}_{0k}$ has non-zero components only in this plane due to the transverse nature of GWs.

\noindent To quantize this system in the NC plane, $x^{j}$ and $p_{j}$ in the above Hamiltonian are replaced by operators ${\hat x}^{j}$ and ${\hat p}_{j}$ satisfying the NC Heisenberg algebra
\begin{eqnarray}
\left[{\hat x}_{i}, {\hat p}_{j}\right] = i\hbar \delta_{ij} \>, \quad 
\left[{\hat x}_{i}, {\hat x}_{j}\right] = i \theta \epsilon_{ij} \>,\quad 
\left[{\hat p}_{i}, {\hat p}_{j}\right] = 0\>.
\label{e9a}
\end{eqnarray}
This can be mapped to the standard $\left( \theta = 0 \right)$ Heisenberg algebra spanned by $X_{i}$ and $P_{j}$ using \cite{mezin}
\begin{eqnarray}
{\hat x}_{i} = X_{i} - \frac{1}{2 \hbar} \theta \epsilon_{ij} P_{j}~,~{\hat p}_{i} = P_{i}.
\label{e9b}
\end{eqnarray}
Now using the traceless property of the GW and substituting eq.$(\ref{e9b})$ in eq.(\ref{e90}), we obtain the NC Hamiltonian in terms of the commutative operators $X_{i}$ and $P_{j}$ 
\begin{equation}
{\hat H} = \frac{ P_{j}{}^{2}}{2m} + \Gamma^j_{0k} X_{j} P_{k} - \frac{\theta }{2 \hbar} \epsilon_{jm} P_{m} P_{k}  \Gamma^j_{0k}
+\mathcal{O}(\Gamma^2)~. 
\label{e112}
\end{equation}
Note that since it has been demonstrated in various formulations of NC gravity \cite{pmas} that the leading NC correction in the gravitational sector is second order, and we are only interested in first order NC effects, we shall use the results of linearized gravity, (unaltered by the NC effect) in the present paper.
Now defining raising and lowering operators
\begin{eqnarray}
X_j &=& \left({\hbar\over 2m\varpi}\right)^{1/2}\left(a_j+a_j^\dagger\right)\>\label{e15a} \\
P_j &=& -i\left({\hbar m\varpi\over 2}\right)^{1/2} \left(a_j-a_j^\dagger\right)
\label{e15}
\end{eqnarray}
where $\varpi$ is determined from the initial uncertainty in either the position or the momentum of the particle, eq.$(\ref{e112})$ 
can be recast as 
\begin{eqnarray}
{\hat H} &=& \frac{\hbar\varpi}{4}\left(2 a_j^\dagger a_j + 1 - a_j^2 - {a_j^\dagger}^2\right) - \frac{i\hbar}{4} \dot h_{jk} \left(a_j a_k - a_j^\dagger a_k^\dagger\right) \nonumber \\
&& + \frac{m \varpi \theta}{8} \epsilon_{jm} {\dot h}_{jk}  \left(a_{m}a_{k}  - a_{m}a_{k}^\dagger + C.C \right)\>
\label{e16}
\end{eqnarray}
where C.C means complex conjugate. Hence the Heisenberg equation of motion of $a_{j}(t)$ reads 
\begin{eqnarray}
\frac{da_j}{dt} = -i\frac{\varpi}{2}(a_j-a^\dagger_j) + \frac{1}{2}\dot{h_{jk}}a^\dagger_k 
+ \frac{i m \varpi \theta}{8 \hbar} \left(\epsilon_{lj} {\dot h}_{lk} + \epsilon_{lk} {\dot h}_{lj}\right)\left(a_{k} - a^\dagger_{k}\right)
\label{e17}
\end{eqnarray}
and that of $a_{j}^{\dagger}(t)$ is the C.C of eq.$(\ref{e17})$. Since the raising and lowering operators must satisfy the commutation relations
\begin{equation}
[a_j(t),a_k(t)] = 0\>,\qquad [a_j(t),a^\dagger_k(t)] = \delta_{jk}\>
\label{e18}
\end{equation}
we write them in terms of $a_j(0)$ and $a_j^{\dagger}(0)$, the free operators at $t=0$, by the time-dependent Bogoliubov transformations
\begin{eqnarray}
a_j(t) &=& u_{jk}(t) a_k(0) + v_{jk}(t)a^\dagger_k(0)\nonumber \\
a_j^\dagger(t) &=& a_k^\dagger(0)\bar u_{kj}(t)  + a_k(0)\bar v_{kj}(t)
\label{e19}
\end{eqnarray}
where the bar denotes the C.C and $u_{jk}$ and $v_{jk}$ (which are $2\times 2$ complex matrices) are the generalized Bogoliubov coefficients which due to eq.(\ref{e18}) satisfies
$uv^{T}=u^{T}v\>,\> u u^\dagger - v v^\dagger = I$.
The initial conditions on $u_{jk}$ and $v_{jk}$ are $u_{jk}(0)= I $ and
$v_{jk}(0) = 0$ since $a_j(t = 0) = a_j(0)$. 

\noindent Defining $\xi = u + v^\dagger$ and $\zeta = u - v^\dagger$, we have (from eq.(\ref{e17}) and its C.C) the following
\begin{eqnarray}
\frac{d \xi_{jk}}{dt} &=& -i\varpi \zeta_{jk} + \frac{{\dot h}_{jl}}{2}\xi_{lk} +\Theta_{jl} \zeta_{lk}\> 
\label{e211a} \\
\frac{d \zeta_{jk}}{dt} &=& -\frac{1}{2}{\dot h}_{jl}\zeta_{lk}\> 
\label{e211b}
\end{eqnarray}
where $\Theta_{jl}$ is the new term reflecting the interplay of noncommutativity with GW
\begin{eqnarray}\Theta_{jl} = \frac{i m \varpi \theta}{4 \hbar}\left({\dot h}_{jm} \epsilon_{ml} - \epsilon_{jm} {\dot h}_{ml}\right)\>.
\label{e21ab}
\end{eqnarray}
Eq(s).~$(\ref{e211a}, \ref{e211b})$ are difficult to solve analytically for general $h_{jk}$. However, our aim is to investigate {\it to what extent spatial noncommutativity affects the interaction of GWs with a spin zero test particle} in the simplest of settings. In this paper, we shall solve eqs.$(\ref{e211a}, \ref{e211b})$ for the case of circularly polarized GWs. 

\noindent In the two-dimensional plane, a circularly polarized GW $h_{jk}$ can be written in terms of the Pauli spin matrices as 
\begin{equation}
h_{jk} \left(t\right) = 2f_{0}\left(\varepsilon_{1}(t)\sigma^1_{jk} + \varepsilon_{3}(t)\sigma^3_{jk}\right) 
= 2f_{0}~\varepsilon_{A}(t)\sigma^A_{jk}~;~A=1,2,3
\label{e13}
\end{equation}
where $2f_{0}$ is the constant amplitude of the GW and $\varepsilon_{1}(t)$ and $\varepsilon_{3}(t)$  representing the two possible 
states of polarization of the GW satisfies the constraint $\varepsilon_{1}^2+\varepsilon_{3}^2 = 1$ for all $t$ and evolve according to 
\begin{equation}
\frac{d\epsilon_{3}(t)}{dt} = \Omega
\epsilon_{1}(t)~,~\frac{d\epsilon_{1}(t)}{dt} = - \Omega\epsilon_{3}(t)
\label{e23}
\end{equation}
where $\Omega$ is a constant.

\noindent We now proceed to solve eqs.$(\ref{e211a})$ and $(\ref{e211b})$. First note that any $2\times 2$ complex matrix $M$ can be written as a linear combination of the Pauli spin matrices and identity matrix as
\begin{equation}
M = \theta_0 I + \theta_A\sigma^A
\label{e23s}
\end{equation} 
where $\theta_0$ and $\theta_A$ are complex numbers. Next, considering $\vec\theta = (\theta_1,\theta_2,\theta_3)$ as being a vector in a three dimensional complex space, it is clear that the polarization states of the GW can also be represented as a vector $\vec\varepsilon$ in this space. Moreover, $\vec\varepsilon$, $\dot{\vec\varepsilon}$ and $\vec\varepsilon\times\dot{\vec\varepsilon}$ are mutually orthogonal and thus form a natural coordinate system for this space. Hence, we make the following ansatz 
\begin{equation}
\zeta = A I + B\vec\varepsilon\cdot\vec\sigma +
       C\frac{\dot{\vec\varepsilon}\cdot\vec\sigma}{\Omega} +
       D~ i\frac{\vec\varepsilon\times
       \dot{\vec\varepsilon}}{\Omega}\cdot\vec\sigma
\label{e24}
\end{equation}
\begin{equation}
\xi =  E I + F\vec\varepsilon\cdot\vec\sigma +
       G\frac{\dot{\vec\varepsilon}\cdot\vec\sigma}{\Omega} +
       H~i\frac{\vec\varepsilon\times
       \dot{\vec\varepsilon}}{\Omega}\cdot\vec\sigma
\label{e25}
\end{equation}
where  the coefficients $A$, $B$, $C$, $D$,$E$,$F$,$G$,$H$ can in general be complex functions. Substituting eqs.(\ref{e24}, \ref{e25}) in eqs.(\ref{e211a}, \ref{e211b}), we get the time evolution equations for the coefficients 
\begin{eqnarray}
\frac{dA}{dt} + f_0\Omega C &=& 0~;~\frac{dB}{dt} - \Omega C-f_0\Omega D = 0 \nonumber \\
\frac{dC}{dt} + \Omega B + f_0\Omega A &=& 0~;~\frac{dD}{dt} - f_0\Omega B = 0\nonumber \\
\frac{dE}{dt} +i\varpi A - f_0\Omega G-i\lambda f_0\Omega B  &=& 0\nonumber \\
\frac{dF}{dt} - \Omega G+i\varpi B + f_0\Omega H-i\lambda f_0\Omega A  &=& 0\nonumber \\
\frac{dG}{dt} + \Omega F +i\varpi C - f_0\Omega E-i\lambda f_0\Omega D &=& 0\nonumber \\
\frac{dH}{dt} +i\varpi D + f_0\Omega F-i\lambda f_0\Omega C  &=& 0
\label{e23a1}
\end{eqnarray}
where $\lambda=\frac{ m \varpi \theta}{ \hbar}$. This way the pair of differential equations (\ref{e211a}, \ref{e211b}) are decomposed into a set coupled differential equations of complex functions. 
Solving them to first order in the GW amplitude with the specified initial conditions, we obtain 
\begin{eqnarray}
A(t)&=&1 - f_0 M_1 ~;~B(t)= M_1 ~;~C(t)=-\left( M_1 + f_0\Omega t\right)~;~D(t)= f_0 M_1 \nonumber\\
E(t)&=&1 + f_0 M_1 + i \left[f_0\left(\lambda M_1 - \frac{\varpi}{\Omega}\right)- \varpi t + f_0 M_2\right] \nonumber\\
F(t)&=& M_1 + i\left[-\frac{\varpi\left(1+M_1\right)}{\Omega} + M_2 + \lambda f_0\Omega t \right]\nonumber\\
G(t)&=& - M_1 + f_0 \Omega t + i\left[-\frac{\varpi\left(1+M_1\right)}{\Omega} - M_3-\frac{f_0 \Omega \varpi t^2}{2}\right]\nonumber\\
H(t)&=& - f_0 M_1 + i f_0\left[\lambda M_1 - \frac{\varpi}{\Omega}-M_3\right]
\label{100z}
\end{eqnarray}
where $M_1$, $M_2$, $M_3$ are given by
\begin{eqnarray}
M_1&=& 1 - \cos \Omega t ~;~M_2= \varpi\left(\frac{\sin\Omega t + \cos\Omega t}{\Omega}-t\right)~;~
M_3= \varpi\left(\frac{\sin\Omega t - \cos\Omega t}{\Omega}-t\right).
\label{110}
\end{eqnarray}
Using these expressions and eqs.(\ref{e24}, \ref{e25}, \ref{e15a}, \ref{e15}, \ref{e19}) yields the expectation values of the components of position and momentum of the particle at any arbitrary time $t$ in terms of the initial expectation values of the position $\left(X_{1}\left( 0 \right), X_{2}\left( 0 \right)\right)$ and momentum $\left(P_{1}\left( 0 \right), P_{2}\left( 0 \right)\right)$. We present the explicit expressions for $\langle X_{1}\left(t\right)\rangle$ and $\langle X_{2}\left(t\right)\rangle$ here.
\begin{eqnarray}
\langle X_{1}\left(t\right)\rangle & = &\left[X_{1}\left( 0 \right) + \frac{P_{1}\left( 0 \right) t}{m}\right]+ M_1\left[\left(\epsilon_3-\epsilon_1-f_0\right)X_{1}\left( 0 \right)+\left(\epsilon_1+\epsilon_3-f_0\right)X_{2}\left( 0 \right)\right]\nonumber\\
&&-f_0 \Omega t \left[\epsilon_1X_{1}\left( 0 \right)+\epsilon_3X_{2}\left( 0 \right)\right]-\frac{\lambda f_0 \Omega t}{m\varpi}\left[\epsilon_3P_{1}\left( 0 \right)+\epsilon_1P_{2}\left( 0 \right)\right]\nonumber\\
&&+\frac{f_0}{m\varpi}\left[-\left(\lambda  M_1 - \frac{\varpi}{\Omega} + M_2\right)P_{1}\left( 0 \right)+\left(\lambda  M_1 -\frac{\varpi}{\Omega} - M_3\right)P_{2}\left( 0 \right)\right]\nonumber\\
&&+\frac{1+M_1}{m\Omega}\left[\left(\epsilon_3+\epsilon_1\right)P_{1}\left( 0 \right)+\left(\epsilon_1-\epsilon_3\right)P_{2}\left( 0 \right)\right]\nonumber\\
&&+\frac{1}{m\varpi}\left[\left(M_3\epsilon_3+M_2\epsilon_1\right)P_{1}\left( 0 \right)-\left(M_2\epsilon_1+M_3\epsilon_3\right)P_{2}\left( 0 \right)\right]\nonumber\\
&&+\frac{ f_0 \Omega t^2}{2 m}\left[\epsilon_1P_{1}\left( 0 \right)-\epsilon_3P_{2}\left( 0 \right)\right]
 \label{x1cir}
\end{eqnarray}
\begin{eqnarray}
\langle X_{2}\left(t\right)\rangle & = &\left[X_{2}\left( 0 \right) + \frac{P_{2}\left( 0 \right) t}{m}\right]+ M_1\left[\left(\epsilon_1+\epsilon_3+f_0\right)X_{1}\left( 0 \right)-\left(\epsilon_3-\epsilon_1+f_0\right)X_{2}\left( 0 \right)\right]\nonumber\\
&&+f_0 \Omega t \left[\epsilon_3X_{1}\left( 0 \right)+\epsilon_1X_{2}\left( 0 \right)\right]-\frac{\lambda f_0 \Omega t}{m\varpi}\left[\epsilon_1P_{1}\left( 0 \right)-\epsilon_3P_{2}\left( 0 \right)\right]\nonumber\\
&&-\frac{f_0}{m\varpi}\left[\left(\lambda  M_1 - \frac{\varpi}{\Omega} - M_3\right)P_{1}\left( 0 \right)+\left(\lambda  M_1 -\frac{\varpi}{\Omega} +M_2\right)P_{2}\left( 0 \right)\right]\nonumber\\
&&+\frac{1+M_1}{m\Omega}\left[\left(\epsilon_1-\epsilon_3\right)P_{1}\left( 0 \right)-\left(\epsilon_1+\epsilon_3\right)P_{2}\left( 0 \right)\right]\nonumber\\
&&+\frac{1}{m\varpi}\left[-\left(M_2\epsilon_1+M_3\epsilon_3\right)P_{1}\left( 0 \right)+\left(M_2\epsilon_3-M_3\epsilon_1\right)P_{2}\left( 0 \right)\right]\nonumber\\
&&-\frac{ f_0 \Omega t^2}{2 m}\left[\epsilon_3P_{1}\left( 0 \right)+\epsilon_1P_{2}\left( 0 \right)\right].
\label{x2cir}
\end{eqnarray}
Reassuringly for $\lambda=0 $ one recovers the classical results for circularly polarized GW in the low-velocity and long-wavelength limit. This shows that our formulation, in particular, the non-trivial construction of coordinate directions using GW polarization vectors and their time derivatives, gives consistent results. Looking at the , in this case the only NC effect comes coupled with the amplitude of circularly polarized GW in the form $\lambda f_0$. Naturally this effect is very small compared to the effect of the passing GW and hence not detectable with our present technology. This is a common feature for both linearly and circularly polarized GW interacting with free paticle. In retrospect, this is not surprising since in the free particle Hamiltonian (\ref{e112}) itself the NC effect appears only in the interaction term. We shall see in the next section that the NC effect is prominent in the case of a harmonic oscillator interacting with a circularly polarized GW.

\section{Harmonic oscillator in noncommutative space interacting with circularly polarized gravitational wave}
\noindent In this section, we move on to the study of the quantum dynamics of a harmonic oscillator in NC space interacting with circularly polarized GWs. The Hamiltonian of the system reads
\begin{equation}
{H} = \frac{1}{2m}\left({p}_{j} 
+ m \Gamma^j_{0k} {x}^{k}\right)^2 + \frac{1}{2} m \varpi^{2} x_{j}^2  \>.
\label{e9h}
\end{equation}
We quantize the system in NC space, following  our prescription given in section 2 and arive at the Hamiltonian
\begin{eqnarray}
{\hat H} = \frac{ P_{j}{}^{2}}{2m} + 
\frac{1}{2} m \varpi^{2} X_{j}{}^{2}+ 
\Gamma^j_{0k} X_{j} P_{k} -
\frac{m \varpi^{2}}{2 \hbar} 
\theta \epsilon_{jm} X^{j} P_{m} 
-\frac{\theta }{2 \hbar} \epsilon_{jm} P_{m} P_{k}  \Gamma^j_{0k} \>. 
\label{e12h}
\end{eqnarray}
The first two terms in the above equation are for the ordinary harmonic oscillator, the third term which is linear in the affine connections shows the effect of the passing GW  on the ordinary harmonic oscillator system, the fourth term is the signature of NC space, a pure NC term linear in the NC parameter and the final term shows the coupling between the GW and spatial noncommutativity\footnote{Since we are dealing with linearized gravity, a term quadratic in $\Gamma$ has been neglected in eq.$(\ref{e12h})$.}.

\noindent Defining raising and lowering operators as in eq.(\ref{e15a}) and eq.(\ref{e15}), but using the oscillator's natural frequency $\omega$ instead of the frequency $\varpi $ fixed by the initial uncertainty of the particle, eq.(\ref{e12h}) can be recast as 
\begin{eqnarray}
{\hat H} = \hbar\omega\left( a_j^\dagger a_j + 1 \right) 
- \frac{i\hbar}{4} \dot h_{jk} 
\left(a_j a_k - a_j^\dagger a_k^\dagger\right)  + \frac{m \omega \theta}{8} \epsilon_{jm} {\dot h}_{jk}  \left(a_{m}a_{k}  - a_{m}a_{k}^\dagger + C.C \right)  -\frac{i}{2} m \omega ^2 \theta\epsilon_{jk}a_{j}^\dagger a_{k}\>.
\label{e16h}
\end{eqnarray}
The time evolution of $a_{j}(t)$ is once again given by the Heisenberg equation of motion :
\begin{eqnarray}
\frac{da_{j}(t)}{dt} &=& -i{\omega}a^j + 
\frac{1}{2}\dot h_{jk}a^\dagger_k - 
\frac{m\omega^{2}\theta}{2\hbar}\epsilon_{jk} a_{k} + \frac{i m \omega \theta}{8 \hbar} \left(\epsilon_{lj} {\dot h}_{lk} 
+ \epsilon_{lk} {\dot h}_{lj}\right)\left(a_{k} - a^\dagger_{k}\right)\>
\label{e17h}
\end{eqnarray}
and that of $a_{j}^{\dagger}(t)$ is the C.C of the above equation. Using the time-dependent Bogoliubov transformations the equations of motions, in terms of the linear combination of Bogoliubov coefficients $\zeta = u - v^\dagger$ and $\xi = u + v^\dagger$, read
\begin{eqnarray}
\frac{d \zeta_{jk}}{dt}=-i\omega \xi_{jk} 
-\frac{1}{2}{\dot h}_{jl}\zeta_{lk} - 
\frac{m \omega^{2}\theta}{2 \hbar}\epsilon_{jl} \zeta_{lk}\> 
\label{e21ah}\\
\frac{d \xi_{jk}}{dt} = -i\omega \zeta_{jk} + 
\frac{1}{2}{\dot h}_{jl}\xi_{lk} +
\Theta_{jl} \zeta_{lk} - \frac{m \omega^{2}\theta}{2 \hbar}
\epsilon_{jl} \xi_{lk}\> 
\label{e21bh} 
\end{eqnarray}
where $\Theta_{jl}$ here is defined as in equation (\ref{e21ab}) with $\varpi$ replaced by $\omega$.

To solve eqs.(\ref{e21ah}, \ref{e21bh}) for the circularly polarized GW (\ref{e13}) we use the ansatz (\ref{e24}, \ref{e25}) to obtain the time evolution of the coefficients 
\begin{eqnarray}
\frac{dA}{dt} +i\omega E+ f_0\Omega C+\Lambda D  &=& 0\nonumber \\
\frac{dB}{dt} +i\omega F- \Omega C-f_0\Omega D  -\Lambda C &=& 0\nonumber \\
\frac{dC}{dt}+i\omega G + \Omega B + f_0\Omega A+\Lambda B &=& 0\nonumber \\
\frac{dD}{dt}+i\omega H - f_0\Omega B-\Lambda A &=& 0\nonumber \\
\frac{dE}{dt} +i\omega A - f_0\Omega G-\frac{2i \Lambda }{\omega}f_0\Omega B+\Lambda H  &=& 0\nonumber \\
\frac{dF}{dt} - \Omega G+i\omega B + f_0\Omega H-\frac{2i \Lambda }{\omega}f_0\Omega A-\Lambda G  &=& 0\nonumber \\
\frac{dG}{dt} + \Omega F +i\omega C - f_0\Omega E-\frac{2i \Lambda }{\omega}f_0\Omega D+\Lambda F &=& 0\nonumber \\
\frac{dH}{dt} +i\omega D + f_0\Omega F-\frac{2i \Lambda }{\omega}f_0\Omega C -\Lambda E &=& 0
\label{eqs}
\end{eqnarray}
where $\Lambda=\frac{m \omega^{2} \theta}{2 \hbar}$.
\noindent Solving these equations to first order in the GW amplitude, with initial conditions on the Bogoliubov coefficients similar to those used in the last section, yields 
\begin{eqnarray}
A(t)&=& 1 - \Lambda K_2 - f_0\Omega K_3+\frac{\Lambda^2-i\omega^2}{\Lambda^2-\omega^2}-i\omega K_1\nonumber\\
B(t)&=&\left(\Omega+\Lambda\right)K_3 +f_0\Omega\left(K_2-\frac{\Lambda}{\Lambda^2-\omega^2}\right)+i\left[-\omega K_4+\frac{\omega\left(\Omega-\Lambda\right)}{\omega^2-\left(\Omega-\Lambda\right)^2}\right]\nonumber\\
C(t)&=&-\left(\Omega+\Lambda\right)K_4 +f_0\Omega\left(K_1+\frac{\Lambda}{\Lambda^2-\omega^2}\right)+\frac{\left(\Omega^2-\Lambda^2\right)}{\omega^2-\left(\Omega-\Lambda\right)^2}-i\omega K_3\nonumber\\
D(t)&=&\Lambda K_1 + f_0\Omega K_4-\frac{f_0\Omega\left(\Omega-\Lambda\right)}{\omega^2-\left(\Omega-\Lambda\right)^2}+\frac{\omega\Lambda\left(1+i\right)}{\Lambda^2-\omega^2} + i\omega K_2 \nonumber\\
E(t)&=&1+f_0\Omega K_3 - \left(\Lambda+i\omega\right)\left(K_1+\frac{\omega}{\Lambda^2-\omega^2}\right)+i\frac{2\Lambda f_0\Omega}{\omega}\left[K_4-\frac{\left(\Omega-\Lambda \right)}{\omega^2-\left(\Omega-\Lambda\right)^2}\right]\nonumber\\
F(t)&=&\left(\Omega+\Lambda\right)K_3-f_0\Omega K_2 +\frac{f_0\Omega\Lambda}{\Lambda^2-\omega^2}+i\left[-\omega K_4+\frac{\omega\left(\Omega-\Lambda\right)}{\omega^2-\left(\Omega-\Lambda\right)^2}+\frac{2\Lambda f_0\Omega}{\omega}\left(K_1+\frac{\omega}{\Lambda^2-\omega^2}\right)\right]\nonumber\\
G(t)&=&-\left(\Omega+\Lambda\right)K_4+f_0\Omega K_1 +\frac{f_0\Omega\omega}{\Lambda^2-\omega^2}+\frac{\left(\Omega^2-\Lambda^2\right)}{\omega^2-\left(\Omega-\Lambda\right)^2}+i\left[-\omega K_3+\frac{2\Lambda f_0\Omega}{\omega}\left(K_2-\frac{\omega}{\Lambda^2-\omega^2}\right)\right]\nonumber\\
H(t)&=&\Lambda K_1 -f_0\Omega\left[K_4-\frac{\left(\Omega-\Lambda \right)}{\omega^2-\left(\Omega-\Lambda\right)^2}\right]+\frac{\omega \Lambda\left(1+i\right)}{\Lambda^2-\omega^2}+i\left[-\omega K_2+\frac{2\Lambda f_0\Omega K_3}{\omega}\right]
\label{100zh}
\end{eqnarray}
where $K_1$, $K_2$, $K_3$, $K_4$ are given by
\begin{eqnarray}
K_1&=& \frac{\left(\sin\omega  t+\cos\omega t\right)\left(\Lambda\sin\Lambda t-\omega\cos\Lambda t\right)}{\Lambda^2-\omega^2} \nonumber\\
K_2&=& \frac{\left(\sin\omega  t+\cos\omega t\right)\left(\Lambda\cos\Lambda t+\omega\sin\Lambda t\right)}{\Lambda^2-\omega^2} \nonumber\\
K_3&=&\frac{\omega\cos\omega t \sin\left(\Omega- \Lambda\right)t-\left(\Omega-\Lambda\right)\sin\omega t \cos\left(\Omega-\Lambda\right)t}{\omega^2-\left(\Omega-\Lambda\right)^2}\nonumber\\
K_4&=&\frac{\omega\sin\omega t \sin\left(\Omega- \Lambda\right)t+\left(\Omega-\Lambda\right)\cos\omega t \cos\left(\Omega-\Lambda\right)t}{\omega^2-\left(\Omega-\Lambda\right)^2}.
\label{100zhh}
\end{eqnarray}
The above expressions together with the set of equations (\ref{e24}, \ref{e25}, \ref{e15a}, \ref{e15}, \ref{e19}) yield the expectation values of the components of position and momentum of the particle at any arbitrary time $t$ in terms of the initial expectation values of the position $\left(X_{1}\left( 0 \right), X_{2}\left( 0 \right)\right)$ and momentum $\left(P_{1}\left( 0 \right), P_{2}\left( 0 \right)\right)$. We provide the explicit expressions for $\langle X_{1}\left(t\right)\rangle$, $\langle X_{2}\left(t\right)\rangle$ and discuss the interesting features they present.
\begin{eqnarray}
\langle X_{1}\left(t\right)\rangle & = & \left[\left(1-\Lambda K_2 + \frac{\Lambda^2}{\Lambda^2-\omega^2}\right)X_{1}\left( 0 \right)+\left(\omega K_1 + \frac{\omega^2}{\Lambda^2-\omega^2}\right)\frac{P_{1}\left( 0 \right) }{m\omega}\right]\nonumber\\
&&+\left(\Omega+\Lambda\right)\left[\left(K_3\epsilon_3-K_4\epsilon_1\right)X_{1}\left( 0 \right)+\left(K_3\epsilon_1+K_4\epsilon_3\right)X_{2}\left( 0 \right)\right]\nonumber\\
&&+f_0\Omega\left[\left(K_3\epsilon_3-K_1\epsilon_1-K_3\right)X_{1}\left( 0 \right)+\left(K_2\epsilon_1+K_1\epsilon_3-2K_4\right)X_{2}\left( 0 \right)\right]\nonumber\\
&&+\left[-\frac{f_0\Omega\omega\left(\epsilon_3+\epsilon_1\right)}{\Lambda^2-\omega^2}X_{1}\left( 0 \right)+\frac{2f_0\Omega\left(\Omega-\Lambda\right)-\left(\Omega^2-\Lambda^2\right)}{\omega^2-\left(\Omega-\Lambda\right)^2}X_{2}\left( 0 \right)\right]\nonumber\\
&&+\frac{1}{m}\left[\left(K_4\epsilon_3+K_3\epsilon_1\right)P_{1}\left( 0 \right)+\left(K_4\epsilon_1-K_3\epsilon_3\right)P_{2}\left( 0 \right)\right]\nonumber\\
&&+\frac{1}{m\omega}\left[-\frac{\omega\left(\Omega-\Lambda\right)\epsilon_3}{\omega^2-\left(\Omega-\Lambda\right)^2}P_{1}\left( 0 \right)+\frac{2\Lambda f_0\Omega}{\omega}\left(\frac{\Lambda \epsilon_3}{\Lambda^2-\omega^2}-K_3\right)P_{2}\left( 0 \right)\right]
\label{x1linear-h}
\end{eqnarray}
\begin{eqnarray}
\langle X_{2}\left(t\right)\rangle & = & \left[\left(1-\Lambda K_2 + \frac{\Lambda^2}{\Lambda^2-\omega^2}\right)X_{2}\left( 0 \right)+\left(\omega K_1 - \frac{\omega^2}{\Lambda^2-\omega^2}\right)\frac{P_{2}\left( 0 \right) }{m\omega}\right]\nonumber\\
&&+\left(\Omega+\Lambda\right)\left[\left(K_3\epsilon_1+K_4\epsilon_3\right)X_{1}\left( 0 \right)-\left(K_3\epsilon_3-K_4\epsilon_1\right)X_{2}\left( 0 \right)\right]\nonumber\\
&&+f_0\Omega\left[\left(K_2\epsilon_1+K_1\epsilon_3+2K_4\right)X_{1}\left( 0 \right)+\left(K_3\epsilon_3-K_1\epsilon_1+K_3\right)X_{2}\left( 0 \right)\right]\nonumber\\
&&+\left[-\frac{2f_0\Omega\left(\Omega-\Lambda\right)+\left(\Omega^2-\Lambda^2\right)}{\omega^2-\left(\Omega-\Lambda\right)^2}X_{1}\left( 0 \right)+\frac{f_0\Omega\omega\left(\epsilon_3+\epsilon_1\right)}{\Lambda^2-\omega^2}X_{2}\left( 0 \right)\right]\nonumber\\
&&+\frac{1}{m}\left[\left(K_4\epsilon_1-K_3\epsilon_3\right)P_{1}\left( 0 \right)-\left(K_4\epsilon_3+K_3\epsilon_1\right)P_{2}\left( 0 \right)\right]\nonumber\\
&&+\frac{1}{m\omega}\left[\frac{2\Lambda f_0\Omega}{\omega}\left(\frac{\Lambda \epsilon_3}{\Lambda^2-\omega^2}-K_3\right)P_{1}\left( 0 \right)+\frac{\omega\left(\Omega-\Lambda\right)\epsilon_3}{\omega^2-\left(\Omega-\Lambda\right)^2}P_{2}\left( 0 \right)\right].
\label{x1linear-hh}
\end{eqnarray}
The solutions clearly reveal the effect of noncommutativity in the response of the harmonic oscillator to the circularly polarized GW. With two frequencies present in the system, one from the circularly polarized GW and other from the harmonic oscillator, it is only natural to expect some resonance behaviour when one frequency approaches the other. But the interesting point is that the spacial noncommutativity introduces an additional frequency $\Lambda$ that is dependent on the NC parameter, the harmonic oscillator frequency and the mass of the test particle into the system. Thus the resonance point is shifted from the expected  value $\Omega = \omega$ by an amount $\Lambda$. In a detector setup, the mass of the test object as well as the natural frequency of the oscillation will be known, therefore the predicted shift in resonance point, if detected, would be able to put stringent bounds on the NC parameter. From equations (\ref{x1linear-h}, \ref{x1linear-hh}) it is evident that there is another resonance point at $\omega  = \Lambda = \frac{m \omega^{2} \theta}{2 \hbar}$ which is realized for a harmonic oscillator with natural frequency $\omega = \frac{2 \hbar}{m \theta}$. However, from the existing upper-bounds on the NC parameter $\theta$ it is clear that for a microscopic test particle, realizing such an oscillator would be difficult.




\section{Conclusions}

\noindent The quantum dynamics of a free particle and a harmonic oscillator interacting with circularly polarized GWs are discussed in a noncommutative setting in the present paper. The results for the free particle exhibits an effect
due to the noncommutativity of space, although it is too small to be identified with our present detection sensitivity. However, the harmonic oscillator case reveals the effect of noncommutativity in an 
interesting way. The results show a resonance phenomena where there is an interplay between the frequency of the GW, natural frequency of the oscillator and frequency set by the noncommutative scale.

\section*{Acknowledgemnet} 
AS acknowledges the supported by DST SERB under Grant No. SR/FTP/PS-208/2012.

\end{document}